\documentclass[aps,prb,12pt,article]{revtex4-1}

\newcommand{\be}{\begin{equation}}
\newcommand{\ee}{\end{equation}}

\usepackage{verbatim}
\usepackage{latexsym,amssymb,float}
\usepackage{setspace}

\usepackage{graphicx}

\usepackage{epsfig}
\usepackage{amsmath}
\usepackage{bm}

\begin{document}

\title{Quantum Engineered Kondo Lattices}
\author{\vspace{0.5cm}Jeremy Figgins$^{1}$, Laila S. Mattos$^{2,3}$, Warren Mar$^{2,4}$, \\
Yi-Ting Chen$^{2,5}$, Hari C. Manoharan$^{2,3,*}$, Dirk K. Morr$^{1,*}$ \\
\vspace{1cm}{\it \normalsize{$^{1}$Department of Physics, University of Illinois at Chicago, Chicago, IL 60607, USA}\\
$^{2}$ Stanford Institute for Materials and Energy Sciences, SLAC National Accelerator Laboratory, Menlo Park, California 94025, USA\\
$^{3}$ Department of Physics, Stanford University, Stanford, California 94305, USA\\
$^{4}$ Department of Electrical Engineering, Stanford University, Stanford, California 94305, USA\\
$^{5}$ Department of Applied Physics, Stanford University, Stanford, California 94305, USA\\ }
\vspace{1cm}
$^\ast$To whom correspondence should be addressed; \texttt{dkmorr@uic.edu}, \texttt{manoharan@stanford.edu}.}

\maketitle
\nopagebreak

\textbf{
Recent advances in atomic manipulation techniques have provided a novel bottom-up approach to investigating the unconventional properties and complex phases of strongly correlated electron materials \cite{Don77,Loh07,Geg08,Sca12,Ste17}.  By engineering artificial condensed matter systems containing tens to thousands of atoms with tailored electronic or magnetic properties \cite{Man00,Moo08,Hir06,Tsu11,Gom12,Slo17,Kim18}, it has become possible to explore how quantum many-body effects---whose existence lies at the heart of strongly correlated materials---emerge as the size of a system is increased from the nanoscale to the mesoscale. Here we investigate both theoretically and experimentally the quantum engineering of nanoscopic Kondo lattices -- Kondo droplets -- that exemplify nanoscopic replicas of heavy-fermion materials. We demonstrate that by changing a droplet's real space geometry, it is possible to not only create coherently coupled Kondo droplets whose properties asymptotically approach those of a quantum-coherent Kondo lattice, but also to markedly increase or decrease the droplet's Kondo temperature. Furthermore we report on the discovery of a new quantum phenomenon -- the Kondo echo -- a signature of droplets containing Kondo holes functioning as direct probes of spatially extended, quantum-coherent Kondo cloud correlations.}

Quantum engineering of finite-size artificial adatom lattices on metallic surfaces has provided a unique approach to explore how variations in the lattice shape or structure affect emergent electronic properties. This has given rise to the creation of Dirac cones in molecular graphene \cite{Gom12} and artificial Lieb lattices \cite{Slo17}, as well as the discovery of novel quantum phenomena, such as quantum mirages in Kondo corrals \cite{Man00,Moo08} or topological superconductivity \cite{Kim18}. Artificial lattices also open a new road to studying the emergence of strong correlation effects through the use of magnetic adatoms, and the ensuing Kondo effect \cite{Kon64}. Such Kondo lattices \cite{Tsu11}, which when rendered coherent, are believed to contain all salient features of heavy fermion materials. They should permit the exploration of how many-body phenomena emerge at the nanoscale, evolve across the mesoscale, and result in the many complex properties of macroscopic systems \cite{Wir16}. Indeed, we find that coherent Kondo droplets can be created with fewer than 50 adatoms, opening a new arena for the exploration of heavy fermion physics at the nanoscale. At the same time, atomic manipulation techniques enable the controlled implementation of defects or vacancies, opening up a new field for studying the interplay between disorder and strong correlation effects. These possibilities might hold the key to greatly advancing our understanding of the many unconventional properties and complex phases of strongly correlated electron materials at the macroscale \cite{Loh07,Geg08}.

\begin{center}
\begin{figure}
\includegraphics[width=5.in]{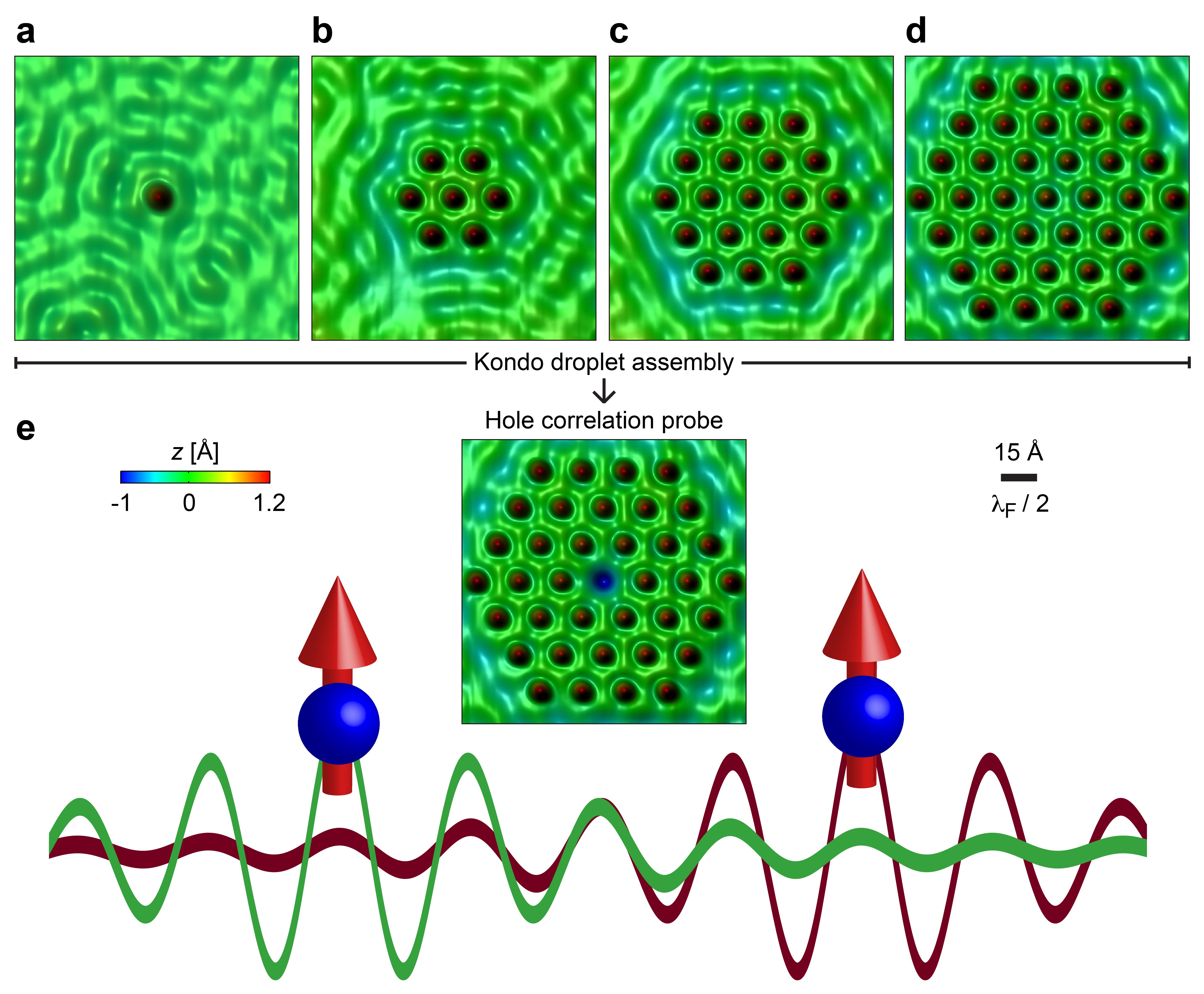}%
 \caption{{\bf Assembly of Kondo droplets using atomic manipulation.} Starting from a single Co atom, {\bf a}, consecutive construction of a Kondo droplet with 1 ({\bf b}), 2 ({\bf c}), and 3 ({\bf d}) rings.  In {\bf e}, a single atom is removed to create a Co vacancy (Kondo hole) in the droplet center, used as a Kondo cloud correlation probe for the entire droplet. Here, the Co adatom distance is $\Delta r_2=4\sqrt{3} a_0 \approx 17.7$ \AA, with $a_0 = 2.55$ \AA\ being the Cu(111) surface nearest-neighbor spacing at $T=4.2$ K. Schematic: Quantum interference of electronic wave functions participating in the Kondo screening of magnetic Co atoms on nearby sites can either enhance or suppress Kondo screening. In the case of constructive interference, a macroscopic Kondo cloud pervades the entire droplet and is detected in the hole site as a Kondo echo.}
 \label{fig:Fig0}
 \end{figure}
 \end{center}
\begin{figure}
\includegraphics[width=3.8in]{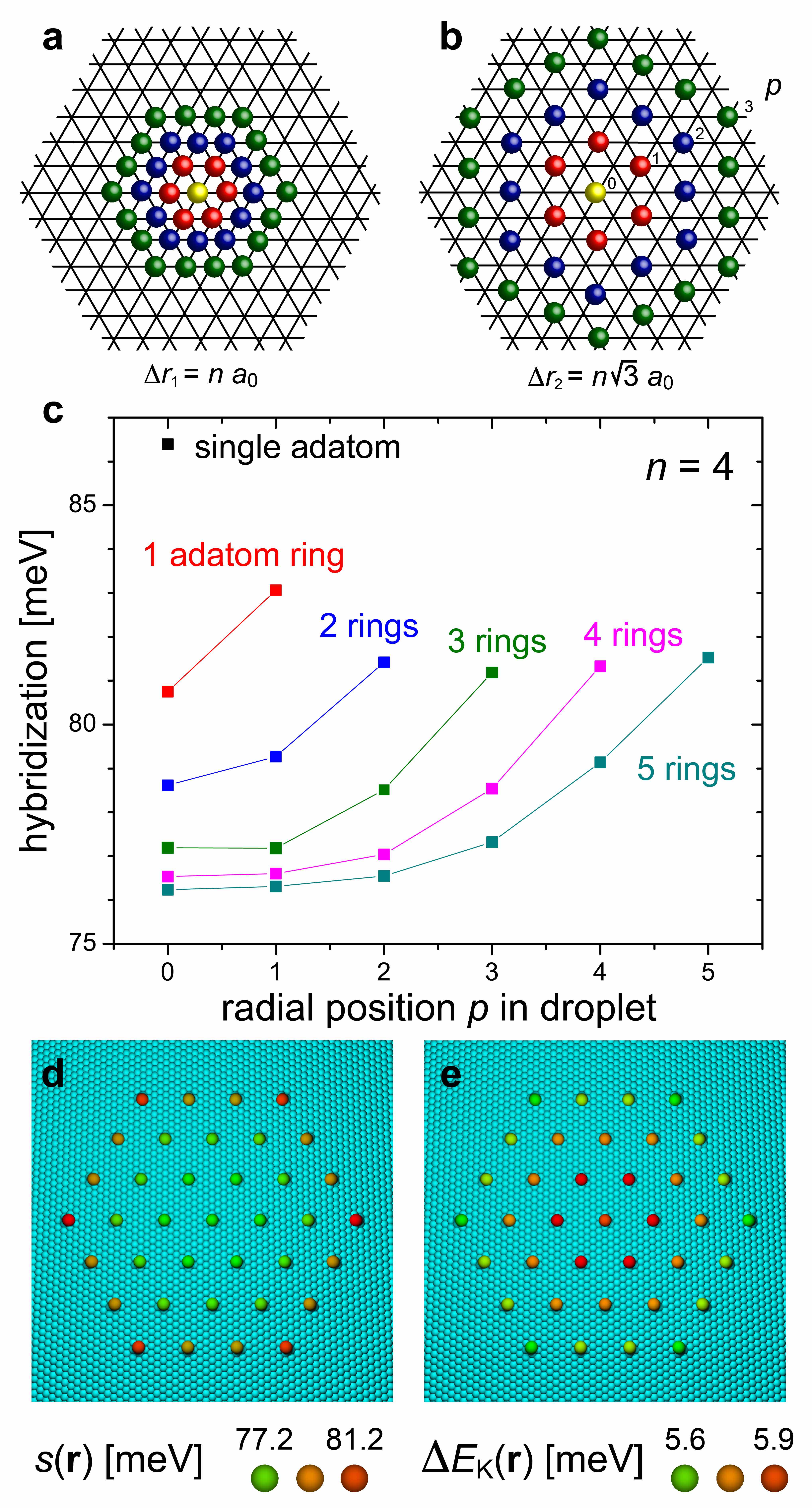}%
 \caption{{\bf Spatially dependent Kondo screening within a droplet.} Spatial structure of Kondo droplets consisting of three rings of magnetic adatoms with distance {\bf a}, $\Delta r_1 = n a_0$ and {\bf b}, $\Delta r_2 = \sqrt{3} n a_0$, with $n$ integer, here shown for $n=1$. For clarity, the rings are shown using different adatom colors. {\bf c}, Hybridization $s({\bf r})$ for different sites in the Kondo droplet shown in {\bf b} with  $\Delta r_2 = 4 \sqrt{3} a_0$ with increasing number of rings of magnetic adatoms. Spatial structure of {\bf d}, $s({\bf r})$ and {\bf e}, $\Delta E_K({\bf r})$ for a Kondo droplet with 3 rings of adatoms and $\Delta r_2 = 4 \sqrt{3} a_0$.}
 \label{fig:Fig1}
 \end{figure}

To study the emergence of correlation effects in Kondo droplets, we use atomic manipulation to arrange magnetic Co adatoms on metallic Cu(111) surfaces \cite{Man00} in the form of highly ordered, hexagonal Kondo droplets (Fig.~\ref{fig:Fig0}{\bf a--d}). After benchmarking the ``intact'' lattices, we study defects in the form of Kondo holes created through ``missing'' Co adatoms (Fig.~\ref{fig:Fig0}{\bf e}). Cu(111) surfaces are uniquely suited to investigate the emergence of many-body effects in Kondo droplets, as Kondo screening arises from the coupling to a surface band of two-dimensional electrons \cite{Man00} -- in which correlation effects are expected to be spatially longer ranged -- rather than to three-dimensional bulk bands. Such Kondo droplets are described by the Kondo Hamiltonian \cite{Don77,Kon64,Col83,Hew93,Si03,Sen04,Oha05,Paul07}
\begin{equation}
{\cal H} = \sum_{{\bf r,r'},\sigma} \left( - t_{{\bf rr'}} - \mu \, \delta_{{\bf rr'}} \right)
c^\dagger_{{\bf r},\sigma} c_{{\bf r'},\sigma} + J {\sum_{{\bf
r}}}' {\bf S}^{K}_{\bf {\bf r}} \cdot {\bf s}^c_{\bf r}  \label{eq:1} \ , 
\end{equation}
where $c^\dagger_{{\bf r},\sigma}$  $(c_{{\bf r},\sigma})$ creates (annihilates) a conduction electron with spin $\sigma$ at site ${\bf r}$ on the Cu surface. Here, $t_{{\bf rr'}}=0.924$ eV is the fermionic hopping element between nearest-neighbor sites in the triangular Cu (111) surface lattice, and $\mu= -5.13$ eV is its chemical potential, yielding a Fermi wavelength of $\lambda_F \approx 11.5 a_0$, where $a_0$ is the Cu lattice constant \cite{Gom12,Moo08}. Moreover, $J>0$ is the Kondo coupling, and ${\bf S}^{K}_{\bf r}$ and ${\bf s}^c_{\bf r}$ are the spin operators of the magnetic Co adatoms and the conduction electron at site ${\bf r}$, respectively. The primed sum runs over the locations of the Co adatoms only. We note that the lattice structure of the Cu(111) surface allows for the creation of two natural types of hexagonal Kondo droplets (which are rotated by $30^\circ$ with respect to each other) with adatom distances of $\Delta r_1 = n a_0$ (see Fig.~\ref{fig:Fig1}{\bf a}) and $\Delta r_2 = n \sqrt{3} a_0$ (see Fig.~\ref{fig:Fig1}{\bf b}) with $n$ being an integer.

To describe the Kondo screening of the Co adatoms by Cu two-dimensional electrons, we employ a large-$N$ expansion \cite{Col83,Hew93,Sen04,Paul07,Read83,Bic87,Aff88}
which has previously been used to successfully explain \cite{Fig10,Morr17} the asymmetric form of the Kondo resonance in the differential conductance, $dI/dV$, measured via scanning tunneling spectroscopy (STS). Here,
${\bf S}^{K}_{\bf r}$ is generalized to $SU(N)$ and represented via
Abrikosov pseudofermions $f^\dagger_{{\bf r},m}, f_{{\bf r},m}$ which obey the
constraint $\sum_{m=1..N} f^\dagger_{{\bf r},m} f_{{\bf r},m}=1$ with $N=2S+1$ being the
spin degeneracy of the magnetic adatom. This constraint is
enforced at every site of the Kondo droplet by means of a Lagrange multiplier $\varepsilon_f({\bf r})$, while
the exchange interaction in Eq.(\ref{eq:1}) is decoupled via a
hybridization field, $s({\bf r})$. For fixed $J$, we then obtain
$\varepsilon_f({\bf r})$ and $s({\bf r})$ on the saddle point level by minimizing the
effective action \cite{Read83} [for details see Supplemental Information (SI) Sec.\ I].
We note that the Ruderman - Kittel - Kasuya - Yosida (RKKY) interaction between magnetic moments \cite{Don77,Hew93,Oha05} for the 2D Cu(111) surface band decays rapidly with increasing distance \cite{Sim11}, and is significantly smaller than $k_B T_K$ for the relevant inter-adatom distances considered below, such that it can be neglected. Finally, consistent with the line shape and width of the Kondo resonance, $\Delta E_K \approx 4.7$ meV, measured experimentally \cite{Man00} for a single Co adatom on a Cu(111) surface, we employ $J=3.82$ eV and $N=4$ (Ref.~\onlinecite{Fig10}).

To investigate the evolution of Kondo screening from a single Kondo impurity to the Kondo lattice, we begin by considering how the droplet's physical properties change with increasing number of rings of Co adatoms with distance $\Delta r_2 = 4 \sqrt{3} a_0$ (see Fig.~\ref{fig:Fig1}{\bf b}). To this end, we present in Fig.~\ref{fig:Fig1}{\bf c} the self-consistently computed hybridization, $s({\bf r})$, for different adatom sites in the Kondo droplet (as denoted in Fig.~\ref{fig:Fig1}{\bf b}) and increasing number of adatom rings. A key physical fact that emerges is that the hybridization in the center of the droplet --- denoted as site $p=0$ --- decreases with increasing number of rings, with the decreasing rate of change between consecutive rings indicating that it approaches an asymptotic value. As the hybridization for a single Kondo impurity is in general larger than that for a Kondo lattice --  essentially since the conduction electrons need to take part in the Kondo screening of multiple magnetic adatoms -- this decrease of $s$ with increasing droplet size reflects the crossover in the droplet's properties from those of a single Kondo atom to those of a Kondo lattice. Moreover, as one moves from the droplet's center to its edge, the hybridization increases, indicating that the properties of the Kondo adatoms along the edge lie between those of the bulk (as exemplified by the droplet's center) and those of a single Kondo impurity. In Figs.~\ref{fig:Fig1}{\bf d} and {\bf e}, we present a spatial plot of $s({\bf r})$ and of the width of the Kondo resonance, $\Delta E_K({\bf r})$ (which is proportional to the Kondo temperature \cite{Hew93}, $T_K$, and extracted from the $dI/dV$ lineshape using a Fano fit \cite{Fan61}; see SI Sec.\ II) at the site of the Co adatoms, respectively, for a Kondo droplet with 3 rings of magnetic adatoms. It is intriguing that these two properties are anticorrelated: while $s({\bf r})$ in the center of the droplet is smaller than at the edge (for the reasons discussed above), $\Delta E_K({\bf r})$ is larger in the center. Indeed, with increasing number of rings, the suppression of the density of states in the Kondo resonance becomes larger and more extended in energy as it evolves into the hybridization gap \cite{Rac18}, thus explaining the spatial structure of the width of the Kondo resonance shown in Fig.~\ref{fig:Fig1}{\bf e}. These results imply that it is possible to study the crossover from a single Kondo impurity to the Kondo lattices not only with increasing droplet size, but also within the same droplet exhibiting bulk properties in its center and properties more similar to isolated impurities along its edges.

 \begin{figure}
\includegraphics[width=5.in]{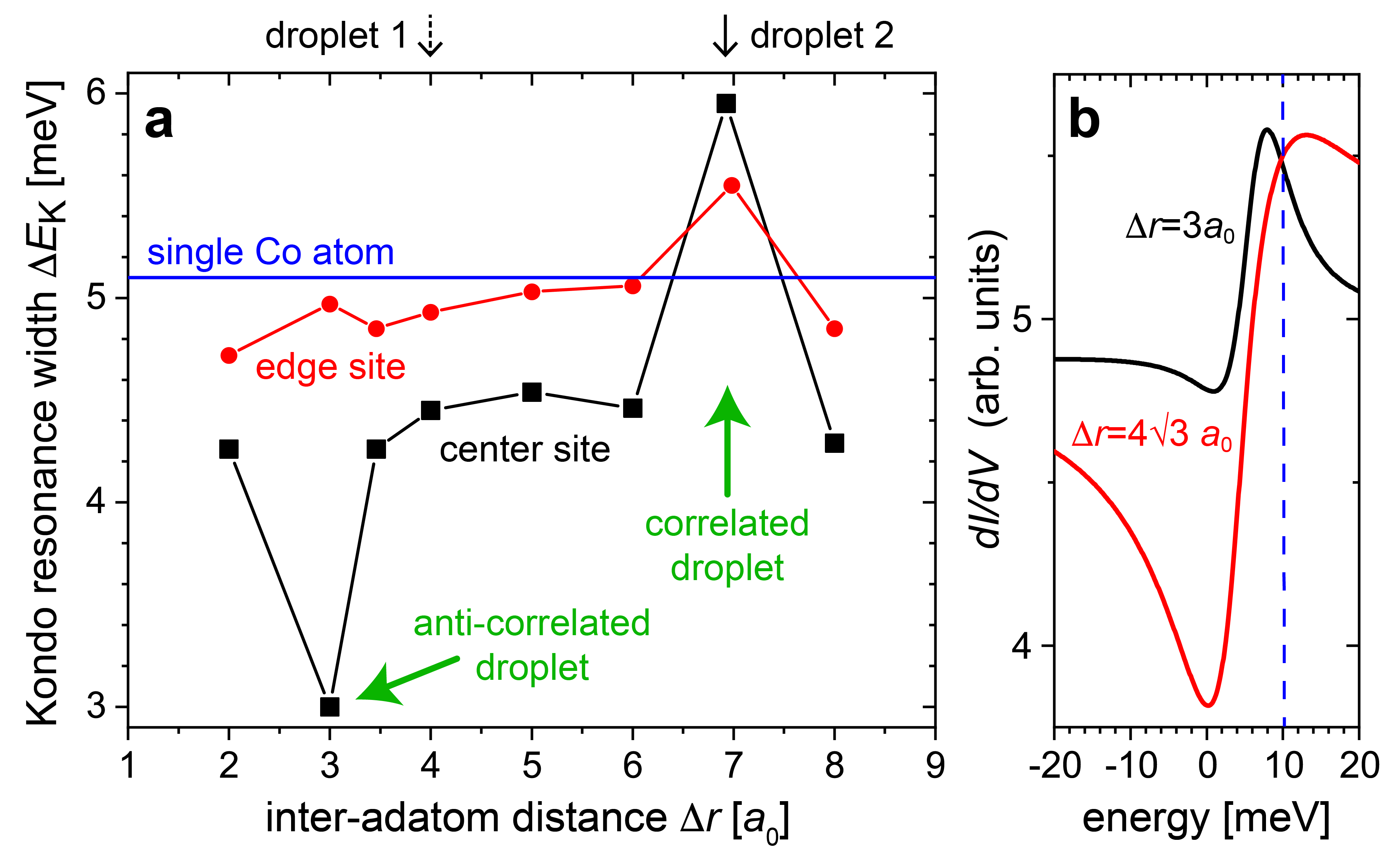}%
 \caption{{\bf Engineering the strength of Kondo screening.} {\bf a}, Dependence of $\Delta E_K({\bf r})$ (a measure for the Kondo temperature) on the inter-adatom distance $\Delta r$ for Kondo droplets consisting of 3 rings. {\bf b}, $dI/dV$ at the center site of a 3-ring Kondo droplet for $\Delta r = 4 \sqrt{3} a_0$ (red curve, maximally correlated droplet) and  $\Delta r = 3 a_0$ (black curve, anti-correlated droplet).  These two lattices correspond to green arrows in {\bf a} contrasting the droplet behavior for constructive (correlated droplet) and destructive (anti-correlated droplet) quantum interference.  At other lattice spacings (see {\bf a}), the droplet is uncorrelated and Kondo screening occurs essentially independently at each decoupled site. The vertical black arrows above {\bf a} indicate experimental droplets investigated which contrast uncorrelated (dashed arrow, droplet 1) and correlated (solid arrow, droplet 2) Kondo droplets.}
 \label{fig:Fig2}
 \end{figure}
The hybridization as well as the width of the Kondo resonance, $\Delta E_K$ (and hence $T_K$), can be controlled in a non-monotonic fashion, by varying the distance between the magnetic adatoms, $\Delta r$. To demonstrate this, we present in Fig.~\ref{fig:Fig2}{\bf a} the width of the Kondo resonance, $\Delta E_K({\bf r})$, as extracted from the theoretical $dI/dV$ lineshape (see SI Sec.\ II), at the center and edge sites of a Kondo droplet with 3 rings for increasing inter-adatom distance. Not only is the dependence of $\Delta E_K$ on $\Delta r$ non-monotonic, but $\Delta E_K({\bf r})$ can be significantly enhanced or suppressed from that of a single Kondo impurity (solid blue line), with the largest enhancement and suppression occurring for $\Delta r=4 \sqrt{3}  a_0$ and $\Delta r=3 a_0$, respectively. The large difference in $\Delta E_K({\bf r})$ is clearly revealed in the $dI/dV$ lineshape at the center sites for these two Kondo droplets, shown in  Fig.~\ref{fig:Fig2}{\bf b}.  This non-monotonic behavior of $\Delta E_K$ arises from the constructive or destructive quantum interference between the Kondo screening clouds associated with each individual magnetic adatom  (Fig.~\ref{fig:Fig0}{\bf e}, and SI Sec.\ III). We note that with increasing droplet size, the width of the Kondo resonance increases and evolves into the hybridization gap of the coherent Kondo lattice \cite{Rac18}. As such, the increased $\Delta E_K({\bf r})$ for $\Delta r_2=4 \sqrt{3} a_0$ is strong evidence for the coherent and constructive coupling of the Kondo screening clouds associated with the individual magnetic adatoms, and hence the creation of a coherent Kondo droplet hosting extended Kondo cloud correlations. On the other hand, for $\Delta r=3 a_0$ the Kondo resonance at the center of the droplet is strongly suppressed, indicating the destructive interference of the Kondo screening clouds. This demonstrates that the creation of a coherent, nanoscale Kondo lattice can be controlled through manipulation of the droplet's geometry and the ensuing quantum interference processes.

\begin{figure}
\includegraphics[width=4.5in]{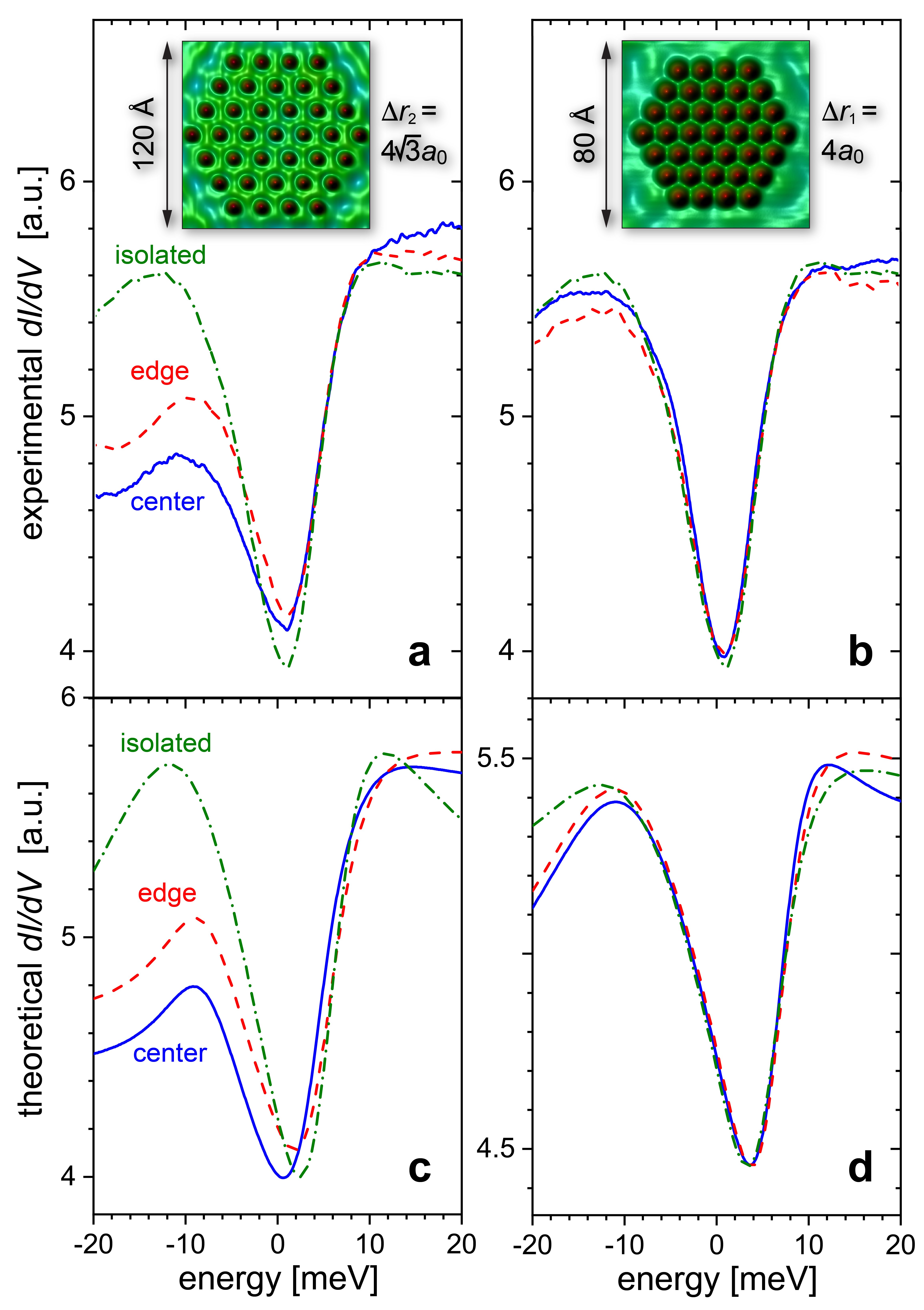}%
 \caption{{\bf Observation of enhanced Kondo screening.} {\bf a}, Experimental $dI/dV$ spectra measured for the Kondo droplets shown in the insets with lattice spacings {\bf a}, $\Delta r_2=4 \sqrt{3} a_0 \approx 17.7$ \AA\ (droplet 2) and {\bf b}, $\Delta r_1=4 a_0 \approx 10.2$ \AA\ (droplet 1) built on the Cu(111) close-packed surface with an exact 6-fold symmetry. {\bf c,d} Theoretical spectral lineshapes for the Kondo droplets shown in the insets of {\bf a,b}, respectively (for detail, see SI Sec.\ IV). [Note that the experimental data show a weak scattering peak around $V=-8$ mV, which arises from the presence of proximal step edges, and is not directly related to the Kondo resonance (this mode was also included in the theoretical fits)].}
 \label{fig:Fig3}
 \end{figure}
While our theoretical results predict the largest difference in the strength of the Kondo screening between droplets with $\Delta r=4 \sqrt{3}  a_0$ and $\Delta r=3 a_0$, we find experimentally, that Co lattices with $n<4$ are insufficiently stable for detailed spectroscopy. On the other hand, for the two lattice types shown in Fig.\ 2{\bf a,b}, the $n=4$ family yields nearest-neighbor magnetic moment distances closest to and straddling  $\lambda_F/2 \sim 15$~\AA, corresponding to the screening limit of $\sim 1$ itinerant electron per spin impurity. Moreover, our theoretical results predict a large difference in correlated behavior between the two types of $n=4$ lattices (Fig.~\ref{fig:Fig2}{\bf a}), contrasting a nearly uncorrelated lattice (dashed vertical black arrow) with a maximally coherent state (solid vertical black arrow). Accordingly, we use atomic manipulation techniques to build two Kondo droplets, each consisting of three rings of Co adatoms with adatom distances of $\Delta r_2=4 \sqrt{3} a_0 \approx 17.7$  \AA\ (droplet 2) and $\Delta r_1=4 a_0 \approx 10.2$ \AA\ (droplet 1), where $a_0=2.55$ \AA\ is the surface nearest neighbour spacing for  Cu(111)  at 4.2 K, as shown in the insets of Figs.~\ref{fig:Fig3}{\bf a} and {\bf b}, respectively. In Figs.~\ref{fig:Fig3}{\bf a} and {\bf b}, we present the experimental $dI/dV$ lineshapes measured via STS at the center and edges of these droplets, and for comparison, that of an isolated Co atom. The $dI/dV$ lineshapes for these two droplets exhibit striking differences: the width of the Kondo resonances in droplet 2 are drastically enhanced over that of the isolated Co atom, with the center showing a larger $\Delta E_K$ than the edge site.  Using a Fano fit \cite{Fan61} to extract the width of the Kondo resonance (SI Sec.\ II), and setting $\Delta E_K = k_B T_K$ (Ref.~\onlinecite{Man00}), we find that the largest Kondo temperature $T_K$ observed is $\sim 72$ K at the droplet center, representing an $\sim 30$\% enhancement over the $T_K \sim 55$ K of an isolated Co adatom.  In stark contrast to this behavior, the corresponding $dI/dV$ lineshapes in droplet 1 are nearly identical to that of an isolated Co atom, and the Kondo temperatures are the same within experimental error.  In Figs.~\ref{fig:Fig3}{\bf c} and {\bf d}, we present the theoretically computed $dI/dV$ lineshapes \cite{Fig10,Mal09,Morr17} (see SI Sec.\ IV), which well reproduce the experimental results shown in Figs.~\ref{fig:Fig3}{\bf a} and {\bf b}. These results confirm the theoretically predicted enhancement of the width of the Kondo resonance for $\Delta r_2=4 \sqrt{3} a_0$ (see Fig.~\ref{fig:Fig2}{\bf a}).

\begin{figure}
\includegraphics[width=4in]{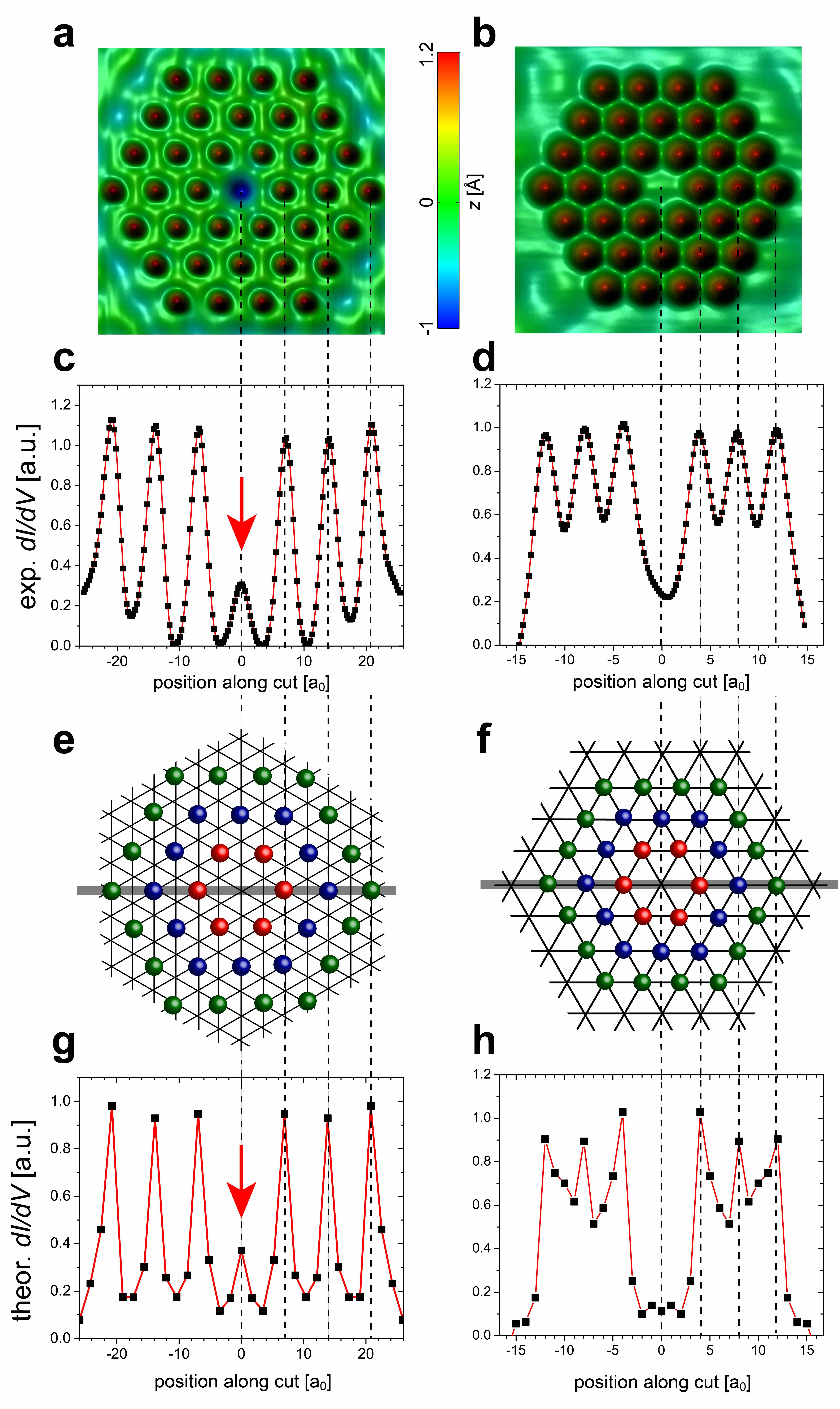}%
\caption{{\bf Emergence of a Kondo echo.} Experimental Kondo droplets for {\bf a}, $\Delta r=4 \sqrt{3} a_0$ (droplet 2) and {\bf b}, $\Delta r=4 a_0$ (droplet 1) with a Kondo hole (vacancy) at the center site. {\bf c,d} Experimental $dI/dV$ at $V=10$ mV acquired along a horizontal cut through the Kondo droplets in {\bf a,b}, respectively. Theoretical Kondo droplets for {\bf e}, $\Delta r=4 \sqrt{3} a_0$ and {\bf f}, $\Delta r=4 a_0$ with a Kondo hole (vacancy) at the center site. {\bf g,h} Theoretical $dI/dV$ at $E=10$ meV along a horizontal cut (gray line) through the Kondo droplets in {\bf e,f} , respectively. The spatial ($x$-axis) scale is the same in panels {\bf a,c,e,g} ({\bf b,d,f,h}).}
 \label{fig:KH}
 \end{figure}
The creation of a coherent Kondo droplet for $\Delta r=4 \sqrt{3} a_0$ leads to another unique effect -- the formation of a {\it Kondo echo}.  When a magnetic adatom is removed from the droplet's center to create a vacancy, or a Kondo hole (as shown in Figs.~\ref{fig:KH}{\bf a,b}), the spectral signature of the Kondo effect can be observed nonlocally at this site as a Kondo echo. To demonstrate this, we present in Fig.~\ref{fig:KH}{\bf c,d} the experimentally measured differential conductance, $dI/dV$, along a line through the center of the coherently-coupled Kondo droplet (see horizontal gray lines in Figs.~\ref{fig:KH}{\bf e,f}) at $E=10$ meV, corresponding approximately to the maximal $dI/dV$ of the Kondo resonance (see dashed line in Fig.~\ref{fig:Fig2}{\bf b}). A comparison of Figs.~\ref{fig:KH}{\bf a} and {\bf c} which both have the same spatial ($x$-axis) scale shows that $dI/dV$ exhibits a peak at the sites of each magnetic adatom. Interestingly enough, however, an additional peak in $dI/dV$ -- the Kondo echo -- occurs at the Kondo hole site in the droplet's center. This Kondo signal represents the correlated signature of the Kondo screening cloud of electrons that is macroscopically coherent over the entire droplet.  As further evidence of this effect, we show that the Kondo droplet with $\Delta r=4 a_0$ (see Fig.~\ref{fig:KH}{\bf b}), in which the Kondo resonances are essentially decoupled from each other, exhibits no Kondo signal at the center vacancy (see Fig.~\ref{fig:KH}{\bf d}). Our theoretical results for $dI/dV$ along the same line for the same Kondo droplets with a Kondo hole at the center (see Figs.~\ref{fig:KH}{\bf e,f}, and SI Sec.\ IV for details) very well reproduce these experimental findings (see Figs.~\ref{fig:KH}{\bf g,h}), and in particular, confirm the existence of a Kondo echo at the vacancy site (Fig.~\ref{fig:KH}{\bf g}). These results thus demonstrate that the existence of a nonlocal Kondo echo reflects the presence of an extended Kondo cloud that is pervasive and coherent over the entire droplet.

The ability to manipulate the strength of Kondo screening, and hence the Kondo temperature, by varying the size, shape, or lattice constant of Kondo droplets opens a new venue for the exploration and quantum engineering of many-body effects, and the many ensuing unconventional properties and phases exhibited by strongly correlated electron materials. Future studies will focus on the possibility to tune the competition between Kondo screening and antiferromagnetic correlations -- which is at the heart of the heavy fermion problem \cite{Don77} -- in the same manner as Kondo screening by itself can be manipulated. The creation of Kondo droplets that balance these two competing phenomena such that the addition of a single magnetic adatom could tip the balance one way or another, presents an exciting opportunity to explore this competition. Furthermore, Kondo screening enables the emergence of unconventional superconductivity in many heavy fermion materials, raising the possibility to create superconducting correlations in Kondo droplets, and manipulate the incipient spatial structure of the superconducting order parameter. The recent development of Josephson scanning tunneling spectroscopy \cite{Ham16,Gra18} would be a suitable tool to probe the emergence of such superconducting correlations on the atomic scale, opening new paths for exploring and manipulating not only the emergence of Kondo lattices and heavy fermion physics, but also that of unconventional superconducting  phases.\\

{\bf Acknowledgments}
This work was supported by the U. S. Department of Energy, Office of Science, Basic Energy Sciences, under Award No. DE-FG02-05ER46225 (JF, DKM), and by the Department of Energy, Office of Science, Basic Energy Sciences, Materials Sciences and Engineering Division, under Contract DE-AC02-76SF00515 (LSM, WM, YTC, HCM).\\

{\bf Methods}
The theoretical results for the local hybridization and Greens functions were obtained using a real space large-$N$ expansion of the Kondo Hamiltonian (for details see SI Sec.~I). The differential conductance, $dI/dV$, was computed assuming simultaneous tunneling into the conduction electron and magnetic $d$-orbitals. The single crystal Cu(111) surface was prepared by repeated cycles of Ar+ ion sputtering and annealing in ultrahigh vacuum.  After cooling the sample to 4.2 K, an electron beam evaporator was used to dose Co to $\sim 4$ atoms/(100 \AA)$^2$ coverage for the Kondo droplet experiments.  The STS tip was then used as an atom manipulator to construct nanoscopic Co lattices comprised of 36 Co atoms each of varying geometry. Using scanning tunneling spectroscopy, the differential conductance, $dI/dV$ was measured. Typical spectroscopic measurement parameters: $R_T = 100$ M$\Omega - 1$ G$\Omega$ at $V = 10$ mV, $V_\text{ac} = 1$ mV rms, $f = 201.4$ Hz.\\

{\bf Author contributions}
D.K.M. and H.C.M. devised the project and wrote the manuscript. J.F. and D.K.M. conducted the theoretical calculations.  L.S.M, W.M, Y.-T.C., and H.C.M. performed experiments and data analysis.

\end{document}


\title{{\huge Quantum Engineered Kondo Lattices}\\[0.5cm]
{\Large Supplemental Information}}

\author{Jeremy Figgins$^{1}$}
\author{Laila S. Mattos$^{2,3}$}
\author{Warren Mar$^{2,4}$}
\author{Yi-Ting Chen$^{2,5}$}
\author{Hari C. Manoharan$^{2,3}$}
\author{Dirk K. Morr$^{1}$}

\affiliation{\vspace{0.5cm}$^{1}$ University of Illinois at Chicago, Chicago, Illinois 60607, USA}
\affiliation{$^{2}$ Stanford Institute for Materials and Energy Sciences, SLAC National Accelerator Laboratory, Menlo Park, California 94025, USA}
\affiliation{$^{3}$ Department of Physics, Stanford University, Stanford, California 94305, USA}
\affiliation{$^{4}$ Department of Electrical Engineering, Stanford University, Stanford, California 94305, USA}
\affiliation{$^{5}$ Department of Applied Physics, Stanford University, Stanford, California 94305, USA}

\maketitle

\section{Theoretical Formalism}

We consider Co adatoms placed on a metallic Cu(111) surface in the form of highly ordered, hexagonal Kondo droplets (see Figs.1 and 2 of the main text).  Kondo screening arises from the coupling of the magnetic adatoms to the two-dimensional surface band. Such a systems is described by the Kondo Hamiltonian
\cite{Kon64,Don77,Col83,Hew93,Si03,Sen04,Oha05,Paul07}
\begin{equation}
{\cal H} = -\sum_{{\bf r,r'},\sigma} t_{{\bf r,r'}}
c^\dagger_{{\bf r},\sigma} c_{{\bf r'},\sigma} + J {\sum_{{\bf
r}}}' {\bf S}^{K}_{\bf {\bf r}} \cdot {\bf s}^c_{\bf r}  \label{eq:1} \ , 
\end{equation}
where $c^\dagger_{{\bf r},\sigma}, c_{{\bf r},\sigma}$ creates (annihilates) a conduction electron with spin $\sigma$ at site ${\bf r}$ on the Cu surface.
Here, $t_{{\bf rr'}}=0.924$ eV is the fermionic hopping element between nearest-neighbor sites in the triangular Cu (111) surface lattice, and $\mu= -5.13$ eV is its chemical potential \cite{Gom12}, yielding a Fermi wavelength of $\lambda_F \approx 11.5 a_0$, where $a_0$ is the Cu lattice constant.
Moreover, $J>0$ is the Kondo coupling, and ${\bf S}^{K}_{\bf r}$ and ${\bf s}^c_{\bf r}$ are the spin
operators of the magnetic (Kondo) atom and the conduction electron
at site ${\bf r}$, respectively. The primed sums run over the locations of the magnetic atoms only. We note that the Ruderman - Kittel - Kasuya - Yosida (RKKY) interaction between magnetic moments \cite{Don77,Hew93,Oha05} for the 2D Cu(111) surface band decays rapidly with increasing distance \cite{Sim11}, and is significantly smaller than $k_B T_K$ for the relevant inter-adatom distances considered in the main text (in particular for the experimental results shown in Figs.4 and 5 of the main text), such that it can be neglected. Finally, to reproduce the line shape and width of the Kondo resonance, $\Delta E_K \approx 4.7$ meV, measured experimentally \cite{Man00} for a single Co adatom on a Cu(111) surface, we employ $J=3.82$ eV [Ref.~\onlinecite{Fig10,Morr17}, and see SI Sec.II]. 

Starting from the Hamiltonian in Eq.(\ref{eq:1}), a systematic large-$N$ expansion \cite{Col83,Hew93,Sen04,Paul07,Read83,Bic87,Mil87,Aff88}
can be achieved by generalizing the spin operators
to $SU(N)$ and representing them using Abrikosov pseudofermions
\begin{equation}
{\bf S}^{K}_{\bf r} =  \sum_{\alpha, \beta} f^\dagger_{{\bf
r},\alpha} {\bm \sigma}_{\alpha, \beta} f_{{\bf r},\beta} \quad
{\bf s}^{c}_{\bf r} =  \sum_{\alpha, \beta} c^\dagger_{{\bf
r},\alpha} {\bm \sigma}_{\alpha, \beta} c_{{\bf r},\beta} \ ,
 \label{eq:slave} 
\end{equation}
where $\alpha,\beta =1,..., N$ and ${\bm \sigma}_{\alpha, \beta}$ is
a vector whose $(N^2-1)$ elements are the generators of $SU(N)$ in
the fundamental representation (each generator is represented by an
$(N \times N)$ matrix with indices $\alpha,\beta$). Here,
$f^\dagger_{{\bf r},\alpha}$ ($f_{{\bf r},\alpha}$) creates (annihilates) a
pseudofermion in the magnetic $d$-orbitals of the Co adatom characterized by the spin quantum number $\alpha$.
To ensure the existence a magnetic moment, one needs to satisfy the constraint that each adatom site is singly-occupied, i.e.,
\begin{equation}
{\hat n}_f({\bf r}) = \sum_\alpha f^\dagger_{{\bf r},\alpha} f_{{\bf r},\alpha} = 1 \ .
\label{eq:const}
\end{equation}
Inserting the representations of Eq.(\ref{eq:slave}) into the
Hamiltonian, Eq.(\ref{eq:1}), yields quartic fermionic interaction
terms. On the mean-field level, we decouple these terms by introducing
the expectation value
\begin{equation}
s({\bf r}) = \frac{J}{2}\sum_{\alpha} \langle f^\dagger_{{\bf
r},\alpha} c_{{\bf r},\alpha} \rangle  \ . 
\label{eq:sr}
\end{equation}
where $s({\bf r})$ describes the local hybridization between the
conduction electron states and the magnetic $f$-electron states. 
$s({\bf r})$ is a measure of the strength of the Kondo screening, with $s({\bf
r})=0$ representing an unscreened magnetic moment at site ${\bf r}$.
The constraint in Eq.(\ref{eq:const}) is enforced on the mean-field level, yielding $ n_f({\bf r}) = \langle {\hat n}_f({\bf r}) \rangle = 1$,  by
adding a Largrange multiplier in the form of the term $\sum_{{\bf r},\alpha} \varepsilon_f({\bf r})
f^\dagger_{{\bf r},\alpha} f_{{\bf r},\alpha}$ to the Hamiltonian in
Eq.(\ref{eq:1}), where $\varepsilon_f({\bf r})$ represents the
on-site energy of the $f$-electron states. The resulting Hamiltonian is
quadratic and can therefore be diagonalized in real space. However, since the lifetime of the conduction and magnetic $d$-orbital states plays a major role
in determining the $dI/dV$ lineshape, we account for them by rewriting the above equation for $s({\bf r})$ [Eq.(\ref{eq:sr}] and the constraint $n_f({\bf r})=1$ in the form
\begin{eqnarray}
s({\bf r}) &=& -\frac{J}{\pi} \int_{-\infty}^{\infty} d\omega \
n_F(\omega) \ {\rm
Im} G_{fc}({\bf r},{\bf r},\omega)  \ ; \nonumber \\
n_f({\bf r}) &=& - \frac{1}{\pi}
\int_{-\infty}^{\infty} d\omega \ n_F(\omega) \ {\rm Im} G_{ff}({\bf
r},{\bf r},\omega)  = 1\ , \label{eq:SC2}
\end{eqnarray}
where
\begin{eqnarray}
{\hat G}_{ff}(\omega) & = &  \left[{\hat g}^{-1}_{ff}(\omega) - {\hat s}
{\hat g}_{cc}^{-1}(\omega) {\hat s}  \right] ^{-1} \ ; \nonumber \\
{\hat G}_{cc}(\omega) & = & \left[{\hat G}_{cc}^{-1}(\omega) -  {\hat s}
{\hat g}_{ff}^{-1}(\omega) {\hat s} \right] ^{-1} \ ; \nonumber \\
{\hat G}_{fc}(\omega) & = & - {\hat g}_{cc}(\omega) {\hat s}
{\hat G}_{ff}(\omega) \ . \label{eq:GF}
\end{eqnarray}
Here, ${\hat G}_{\alpha \beta}(\omega) \ (\alpha,\beta=c,f)$ are Greens function matrices in real space with $G_{\alpha \beta}({\bf r},{\bf r},\omega)$ being the $(r,r)$ element of the matrix ${\hat G}_{\alpha \beta}$. ${\hat g}_{ff}$ and ${\hat g}_{cc}$ are the unhybridized Greens function matrices, and the spin degeneracy has been accounted for by dropping the spin index in the matrices. The hybridization matrix ${\hat s}$ is only non-zero at the sites where a magnetic adatom is located. A finite lifetime, $\tau_{c,f}$, of the conduction and $d$-orbital states is then introduced in the above Greens functions via the scattering rate $\Gamma_{c,f} = \hbar/ \tau_{c,f}$ yielding
\begin{eqnarray}
{\hat g}_{ff}({\bf r},{\bf r},\omega) &=& \left[ \omega - \varepsilon_f({\bf r}) + i \Gamma_f \right]^{-1} \\
{\hat g}_{cc}({\bf k},\omega) &=& \left[ \omega - \varepsilon_{\bf k} + i \Gamma_c \right]^{-1} \ . \label{eq:GF0}
\end{eqnarray}
with ${\hat g}_{cc}({\bf r},{\bf r},\omega)$ being obtained from ${\hat g}_{cc}({\bf k},\omega)$ via Fourier transform. To reproduce the experimentally measured $dI/dV$ lineshapes shown in Fig.4 of the main text, we used $\Gamma_{f}=4$ meV and $\Gamma_{c}=65$ meV.
Eqs.(\ref{eq:SC2}) - (\ref{eq:GF0}) are a closed set of equations that can now be self-consistently solved to obtain the local hybridizations, $s({\bf
r})$ and $f$-electron energies $\varepsilon_f({\bf r})$. We note that for a single Kondo impurity, this
formalism is identical to the saddle-point approximation of the
path-integral approach of Ref.~\cite{Read83} in the large-$N$ approximation which becomes exact in the limit $N \rightarrow \infty$.

\section{Differential Conductance $dI/dV$, the Fano-fit, and the width of the Kondo resonance}

To compute the differential conductance, $dI/dV$
\cite{Mal09,Fig10,Wol10,Morr17}, measured in STS experiments on Kondo systems \cite{Mad01} and heavy fermion materials
\cite{Sch09,Ayn10,Ern11}, we define the spinor $\Psi^\dagger_{{\bf
k},\alpha}=(c^\dagger_{{\bf k},\alpha},f^\dagger_{{\bf k},\alpha})$ and the Green's function
matrix ${\hat G}_\alpha({\bf k},\tau)=-\langle T_\tau \Psi_{{\bf
k},\alpha}(\tau)\Psi^\dagger_{{\bf k},\alpha}(0) \rangle$. With $t_c$ and $t_f$
being
the amplitudes for electronic tunneling from the tip into the conduction electron bands of the Cu(111) surface and the magnetic $d$-orbitals of the Co adatom at site ${\bf r}$,
respectively, one obtains \cite{Fig10}
\begin{eqnarray}
\frac{dI({\bf r},\omega)}{dV} = - \frac{e}{\hbar} N_t
\sum_\alpha \sum_{i,j=1}^2 \left[  {\hat t} \, {\rm Im} {\hat G}_\alpha({\bf
r,r},\omega) \, {\hat t} \right]_{ij} \label{eq:dIdV}
\end{eqnarray}
where ${\hat t} =
\begin{pmatrix}
t_c& 0 \\
0 & t_f
\end{pmatrix}$, and $N_t$ is the STS
tip's density of states, which is taken to be constant. In the weak-tunneling limit, $t_c,t_f
\rightarrow 0$, we obtain
\begin{align}
\frac{dI({\bf r},V)}{dV} =& \frac{2 \pi e}{\hbar} N_t \left[t_c^2
N_{c}({\bf r}, eV) + t_f^2 N_{f}({\bf r}, eV)   + 2 t_c t_f N_{cf}({\bf r}, eV) \right]
\label{eq:dIdV}
\end{align}
where
\begin{align}
N_{c}({\bf r},\omega) =& - \frac{1} {\pi} {\rm Im} G_{cc}({\bf r},{\bf r},\omega) \nonumber \\
N_{f}({\bf r},\omega) =& - \frac{1} {\pi} {\rm Im} G_{ff}({\bf r},{\bf r},\omega) \nonumber \\
N_{cf}({\bf r},\omega) =& - \frac{1} {\pi} {\rm Im} G_{cf}({\bf r},{\bf r},\omega) \\
\end{align}
with $N_c$ and $N_f$ being the density of states of the
conduction and $f$-electron ($d$-orbital) states, respectively. To reproduce the line shape and width of the Kondo resonance, $\Delta E_K \approx 4.7$ meV, measured experimentally \cite{Man00} for a single Co adatom on a Cu(111) surface, we employ $J=3.82$ eV and $N=4$ [Ref.~\onlinecite{Fig10,Morr17}]. Moreover, the theoretical fits to the experimental $dI/dV$ results in Figs.4 and 5 of the main text were obtained with the same set of parameters, $J, \Gamma_c$, and $\Gamma_f$. Note that the experimental data in Fig.4{\bf a,b} of the main text show a weak scattering peak around $V=-8$ mV, which arises from the presence of proximal step edges, and is not directly related to the Kondo resonance. This mode, in the form of a Lorentzian peak centered at $E=-8$ meV with half-width $\Delta E= 5.75$ meV as well as a sloping background were added to the theoretically computed $dI/dV$ lineshape, yielding the results shown in Figs.4{\bf c,d}.

To extract the width, $\Delta E_K$ of the Kondo resonances (shown in Figs.2{\bf e} and 3{\bf a} of the main text), we have fitted our theoretically obtained $dI/dV$ lineshapes (shown, for example, in Fig.3{\bf b} of the main text) by the Fano formula \cite{Fano61}
\be
\frac{dI({\bf r},V)}{dV} = y_0 + c V + B \frac{\left( \frac{V-\alpha}{\Delta E_K}+q\right)^2}{\left( \frac{V-\alpha}{\Delta E_K}\right)^2+1}
\ee
where $\Delta E_K$ measures the width of the Kondo resonance, $q$ its asymmetry, and $c V$ represents a sloping background. We find that changing the tunneling ratio, $t_f/t_c$, changes the values for $\alpha$ and $q$, but has only a weak effect on the width of the resonance $\Delta E_K$.

\section{Quantum Interference and Kondo screening}

To demonstrate how quantum interference of the Kondo screening clouds associated with different magnetic adatoms can give rise to variations in the strength of the Kondo screening, as evidenced by the magnitude of the hybridization, $s$, or the width of the Kondo resonance, $\Delta E_K$ (as shown in Figs.2 and 3 of the main text) we begin by
\begin{figure}[h]
\includegraphics[width=7in]{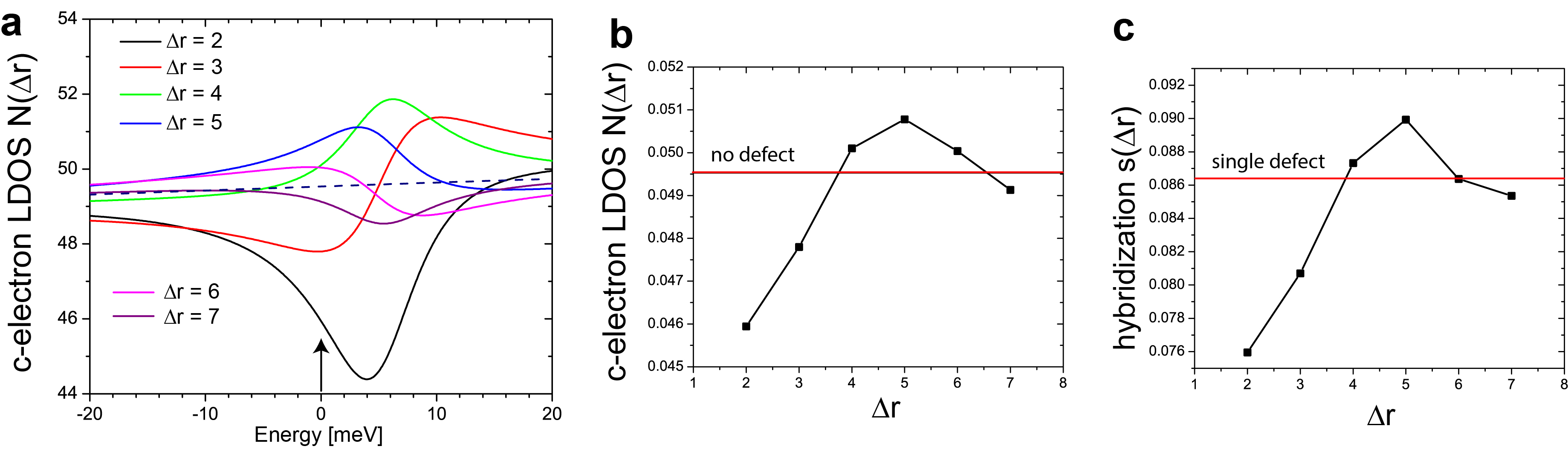}
\caption{{\bf a} Theoretical LDOS of the conduction $c$-electrons at several distances $\Delta$r from a single Co atom on a Cu(111) surface (same parameters as in the main text). {\bf b} c-electron LDOS at $E=0$ as a function of $\Delta$r. The red line represents the LDOS of the unperturbed Cu(111) surface. {\bf c} Theoretically computed hybridization of 2 Co atoms as a function of inter-adatom distance, $\Delta$r. The red line shows the hybridization for a single Co atom.} \label{fig:Fig2}
\end{figure}
considering the effects of a single, Kondo-screened Co atom on the electronic structure in its vicinity.
To this end, we present in Fig.~\ref{fig:Fig2}{\bf a} the local density of states (LDOS) of the conduction $c$-electrons for several distances $\Delta$r from a single, Kondo screened Co adatom located an a Cu(111) surface. The non-monotonic dependence of the LDOS on $\Delta$r arises from $2k_Fr$ scattering of the screening conduction electrons.  Let us now consider the effects on the hybridization when a second Co adatom is added to the system  by noting that the magnitude of the hybridization is in general determined by the value of the conduction electron LDOS -- the screening band -- near the Fermi energy. This LDOS at $E=0$ as a function of $\Delta$r is shown in Fig.~\ref{fig:Fig2}{\bf b}. When a second Co adatom is now placed at a distance $\Delta$r from the first impurity where the LDOS is smaller (greater) than in the unperturbed metal, one expects that the resulting hybridization (which by symmetry is the same for both Co adatoms) is also smaller (greater) than that of a single Co adatom, which is placed on the unperturbed Cu surface. This expectation is confirmed by the explicit calculation of the hybridization of a 2 Co adatom system, as a function of inter-adatom distance, $\Delta$r, shown in Fig.~\ref{fig:Fig2}{\bf c}, which shows the same non-monotonic dependence on $\Delta$r as the LDOS in Fig.~\ref{fig:Fig2}{\bf b}. In Kondo droplets consisting of multiple Co adatoms, it is the same effect as discussed above for the 2 Co adatom system that leads to the non-monotonic dependence of the hybridization on $\Delta$r: it is the quantum interference of the Kondo screening clouds associated with each of the Co adatoms that determines the resulting spatial distribution of the hybridization.

\section{Position dependence of $dI/dV$ in closed loop STS experiments}

To compare our theoretical calculations for the spatial dependence of the differential conductance to the experimental results, shown in Figs.5{\bf c,d} of the main text, we note that the STS experiments are performed in closed loop mode. This implies, that at every position, the height of the tip from the sample is adjusted to keep the total current
\be
I({\bf r}) =  \int_0^{V_s} dV \frac{dI({\bf r},V)}{dV} = I_0
\label{eq:I0}
\ee
for a given set point voltage, $V_S$, constant. A change in height, however, implies that the tunneling amplitudes become position dependent, and that therefore Eq.(\ref{eq:dIdV}) needs to be generalized to
\begin{align}
\frac{dI({\bf r},V)}{dV} =& \frac{2 \pi e}{\hbar} t_c^2({\bf r}) N_t \left[ N_{c}({\bf r}, V) + \left(\frac{t_f({\bf r})}{t_c({\bf r})}\right)^2 N_{f}({\bf r}, V)   + 2 \frac{t_f({\bf r})}{t_c({\bf r})} N_{cf}({\bf r}, V) \right]
\label{eq:dIdVr}
\end{align}
Inserting Eq.(\ref{eq:dIdVr}) into Eq.(\ref{eq:I0}) then yields
\begin{align}
t_c^2({\bf r}) = I_0 \left\{ \frac{2 \pi e}{\hbar}  N_t \int_0^{V_s} dV  \left[ N_{c}({\bf r}, V) + \left(\frac{t_f({\bf r})}{t_c({\bf r})}\right)^2 N_{f}({\bf r}, V)   + 2 \frac{t_f({\bf r})}{t_c({\bf r})} N_{cf}({\bf r}, V) \right] \right\}^{-1}
\end{align}
and hence
\begin{align}
\frac{dI({\bf r},V)}{dV} =& I_0 \frac{ N_{c}({\bf r}, V) + \left(\frac{t_f({\bf r})}{t_c({\bf r})}\right)^2 N_{f}({\bf r}, V)   + 2 \frac{t_f({\bf r})}{t_c({\bf r})} N_{cf}({\bf r}, V) }{\int_0^{V_s} dV \left[ N_{c}({\bf r}, V) + \left(\frac{t_f({\bf r})}{t_c({\bf r})}\right)^2 N_{f}({\bf r}, V)   + 2 \frac{t_f({\bf r})}{t_c({\bf r})} N_{cf}({\bf r}, V) \right]}
\label{eq:dIdVr1}
\end{align}
It is this ``normalized" differential conductance, which is shown in Figs.5{\bf g,h} of the main text and accounts for the experimental closed loop mode that needs to be compared to the experimentally measured $dI/dV$ linecuts through a Kondo hole droplet, shown in Figs.5{\bf c,d} of the main text. For our theoretical results, we used $t_f/t_c=0.03$ at the sites of the Co atoms, and $t_f=0$ otherwise.\\

\begin{figure}[h]
\includegraphics[width=8.5cm]{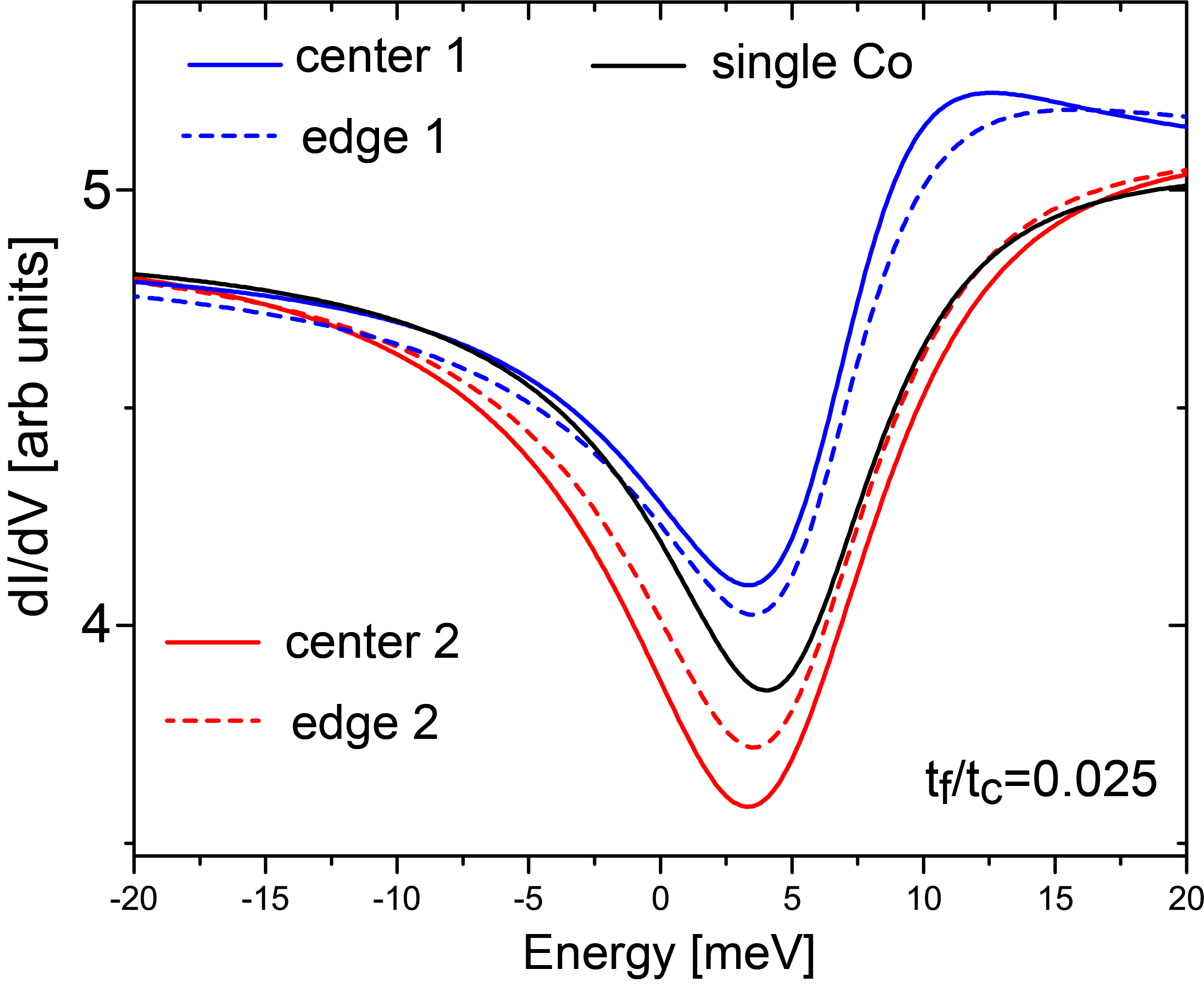}
\caption{Theoretical $dI/dV$ curves with $t_f/t_c=0.025$ for center and edge sites for droplet 1 (1) and droplet 2 (2), as well as for a single, isolated Co adatom.} \label{fig:Fig1}
\end{figure}

We note that when the height of the tip is varied, both $t_c({\bf r})$ and $t_f({\bf r})$ can in general change, which can lead to variations in $t_f/t_c$ not only between the experimentally studied droplets, but also within the same droplet. From a comparison of our theoretical results with the experimental $dI/dV$ lineshapes shown in Figs.4{\bf a,b} of the main text for the Kondo droplets with $\Delta r_2=4 \sqrt{3} a_0$ (droplet 2) and $\Delta r_1=4 a_0$ (droplet 1), respectively, we find that $t_f/t_c$ is slightly larger in droplet 2 than in droplet 1. Specifically, we obtain for droplet 1 $t_f/t_c=0.025$ both at the center and edge sites, while for droplet 2 we obtain $t_f/t_c=0.0327$ at the center site, and $t_f/t_c=0.03$ at the edge site. Finally, for an isolated Co atom, we obtain $t_f/t_c=0.0275$. To show that these small changes in $t_f/t_c$ are related to changes in the tip height (due to the closed loop mode in which the STS experiments are performed) between the different spatial locations for which $dI/dV$ lineshapes are shown in Fig.4 of the main text, we computed the $dI/dV$ lineshapes for all spatial positions with the same value of $t_f/t_c$, as shown in Fig.~\ref{fig:Fig1}. Given the experimentally used set point voltage of $V_s=10$ meV, it immediately follows that if $t_f/t_c$ were constant, that then the total current [see Eq.(\ref{eq:I0})] flowing through the STS tip were larger for droplet 1 than for an isolated Co atom, and larger for an isolated Co atom than for droplet 2. This implies that in order to measure the same total current in the droplets and for an isolated Co atom, the tip height needs to be the smallest for droplet 2, followed by the isolated Co atom and then droplet 1. Given our analysis above, this suggests that the ratio $t_f/t_c$ systematically increases with decreasing tip height, likely due to the increasing overlap of the tip orbitals with the Co d-orbitals.
